\documentclass[11pt]{article}
\usepackage[utf8]{inputenc}
\usepackage[a4paper, margin=2cm]{geometry}
\usepackage{multicol}
\usepackage{graphicx}
\usepackage{authblk}
\usepackage{tikz}
\usepackage{amsmath, amsfonts, amssymb}
\usepackage{xcolor}
\usepackage{hyperref}
\usepackage{caption}
\usepackage[all]{hypcap}
\usepackage{algorithm}
\usepackage{tabularx}
\usepackage[export]{adjustbox}

\usepackage{booktabs}
\usepackage{siunitx}
\usepackage{multirow}
\usepackage{makecell}
\allowdisplaybreaks

\begin{document}
\title{\bf Bounding interventional queries from \\ generalized incomplete contingency tables}

\author[1]{Ivano Lodato\thanks{\href{mailto:ivano.lodato@allos.ai}{ivano.lodato@allos.ai}}}
\author[1,2]{Aditya V. Iyer \thanks{\href{mailto:aditya.iyer@allos.ai}{aditya.iyer@allos.ai}}}
\author[1]{Isaac Z. To\thanks{\href{mailto:isaac.to@allos.ai}{isaac.to@allos.ai}}}
\affil[1]{Allos AI Limited, Hatton Garden, London, EC1 N8LE, UK}
\affil[2]{Clarendon Laboratory, Department of Physics, University of Oxford, Oxford OX1 3PU, UK}
\date{\today}

\maketitle
\begin{abstract}
We introduce a method for evaluating interventional queries and Average Treatment Effects (ATEs) in the presence of generalized incomplete contingency tables (GICTs), contingency tables containing a full row of random (sampling) zeros, rendering some conditional probabilities undefined. Rather than discarding such entries or imputing missing values, we model the unknown probabilities as free parameters and derive symbolic expressions for the queries that incorporate them. By extremizing these expressions over all values consistent with basic probability constraints and the support of all variables, we obtain sharp bounds for the query of interest under weak assumptions of small missing frequencies. These bounds provide a formal quantification of the uncertainty induced by the generalized incompleteness of the contingency table and ensure that the true value of the query will always lie within them. The framework applies independently of the missingness mechanism and offers a conservative yet rigorous approach to causal inference under random data gaps.
\end{abstract}
 \section{Introduction}
The problem of missing data is ubiquitous in real-world data analysis. It can arise due to imperfect experiments (e.g. noncompliance), ethical constraints (e.g. experimental treatments for elderly individuals), design limitations (e.g. difficulty in meeting inclusion criteria in a region), or even physical impossibilities (e.g., the absence of data on three-legged humans). The problem of missingness in datasets has been widely studied in the literature: a variety of statistical methods exist to impute the missing variable(s) values (for reference, see \cite{Kang:13,review2}) under (often restrictive) hypotheses on the mechanism of missingness, in our opinion best explained and exemplified in \cite{Pearl:2020}(Pearl) (see entries $m_{1,2}$ in Table \ref{t0} for example).\\
In general, however, even if a dataset does not contain any missing values, it is still possible that the contingency table contains zeroes. Those are typically divided into two categories: structural/fixed zeros and random/sampling zeros. The first type groups variables that are logically or physically constrained to be zero (e.g. no data exist on a person before their birth or on three-legged man); structural zeros can be cut out from the contingency table as they have no degrees of freedom associated \cite{Baker:85} (for an exhaustive description, see also \cite{smith:90}). Random zeros, instead, are due to small sample sizes or, conversely, to rare events. The expected missing counts are small, at least compared to the observed counts, which breaks down asymptotic (large frequencies) approximations. Nevertheless, under simplifying assumptions on the mechanism of missingness, missing frequencies can be computed using maximum likelihood estimation (see for instance \cite{Dempster:18, brown:83, Brzezińska:15} and for a recent general exposition \cite{fienberg:18})

In this paper, we will take a much more conservative approach and avoid using specific models for the sampling distributions, as well as completely bypass the estimation of the missing values. \\
Instead, we will first assume that random zeros can be modeled as actual zeros, as long as the probability associated with the missing frequency is well-defined (similar to the perspective taken in \cite{Consonni:07}, though with different goals). Secondly, we will propose a way to handle contingency tables containing an entire row of random zeros, leading to ill-defined probabilities. We call this a generalized incomplete contingency table (GICT), the object central to our analysis.
 \section{Generalized incomplete contingency tables}
To illustrate the problem at hand clearly, let us consider the dataset shown in Table \ref{t0} which tracks Age, Gender, and Obesity status for 10 patients. Now, depending on the query of interest, the dataset and, more importantly, the corresponding contingency table may contain missing entries.
Let us consider first a simple example: for the query $P({\rm Age}\in[14,15]|{\rm Gender}\in [0,1])$, the dataset and the contingency table do not suffer from any type of missingness.\\
For a different query, e.g. $P({\rm Obesity}=1|{\rm Gender}=1 ,{\rm Age}=14 )$, the contingency table is incomplete and contains a random zero $Z_0$. Since the complementary probability \\ $P({\rm Obesity}=0|{\rm Age}=14 , {\rm Gender}=1)=1$ is well defined, we can very conservatively assign the actual value zero to the random zero $Z_0$.\\
Finally, consider the query $P({\rm Obesity}=[0,1]|{\rm Gender}=1, {\rm Age}=16)$: the contingency table is now ``generalized incomplete", with a full row of zero values \footnote{We exclude the column ``missing" from the contingency table, since it is relevant to the discussion on dataset missingness but not in this discussion on random zeros in the contingency table. }. This time, we cannot naively impute the value 0 to $Z_{1,2}$, since it leads to inconsistencies, violations of basic probabilistic constraints, and introduces numerical errors into downstream analysis. In fact:\\
\begin{align}
 &P({\rm Obesity}=0,1|{\rm Gender}=1, {\rm Age}=16) = \frac{P({\rm Obesity}=0,1,{\rm Gender}=1, {\rm Age}=16)}{P({\rm Gender}=1, {\rm Age}=16)}=\frac{Z_{1,2}}{Z_1+Z_2}
    \nonumber\\
    &\qquad \quad \sum_{o=0,1}P({\rm Obesity}=o|{\rm Gender}=1, {\rm Age}=16)= \frac{Z_1}{Z_1+Z_2} +\frac{Z_2}{Z_1+Z_2}=\frac{0}{0} \;\;(\neq 1)
    \nonumber
\end{align}
\begin{center}
\begin{tabularx}{\textwidth}{*{2}{>{\centering\arraybackslash}X}}
   \centering
\begin{tabular}[b]{c|ccc}
\hline
N &Age & Gender & Obesity \\
\hline
1 &16 & 0 & 1\\
2 &16 & 0 & 0\\
3 &15 & 0 & $m_1$\\
4 &15 & 0 & 1\\
5 &15 & 0 & 1\\
6 &15 & 1 & $m_2$\\
7 &14 & 0 & 1\\
8 &14 & 0 & 0\\
9 &14 & 0 & 1\\
10 &14 & 1 & 0\\
\hline
\end{tabular}
\captionof{table}{A dataset with 10 entries with Gender $[0,1]$, Age $[14,15,16]$ and Obesity $[0,1]$. With $m_{1,2}$ we indicated the missing data}
\label{t0}
& 
\begin{tabular}[b]{@{}lcccc@{}} 
\toprule 
& & \multicolumn{3}{c}{Obesity} \\ 
\cmidrule{3-5} 
\multirow{-2.5}{*}{\makecell{Age}} & \multirow{-2.5}{*}{\makecell{Gender}} & 0 & 1 & {\rm missing}\\ 
\midrule 
14 & 0 & 1 & 2 & 0 \\
& 1 & 1 & $Z_0$ & 0  \\\addlinespace
15 & 0 &  1 & 2 & $m_1$\\
& 1 & 1 & 0 & $m_2$ \\\addlinespace
16 & 0 & 1 & 1 & 0 \\
& 1 & $Z_1$ & $Z_2$ & 0 \\\addlinespace
\bottomrule 
\end{tabular} 
\captionof{table}{Contingency table obtained from dataset in Table \ref{t0}. $Z_{1,2,3}$ are the random/sampling zeros}
\label{t2}
\end{tabularx}
 \end{center} 
There are ways to circumvent the problem of having generalized random zeros in a contingency table: the easiest is to exclude the fully missing rows from the contingency table and restrict the analysis to the subset of feature values that are empirically represented. While straightforward, this strategy is not always advantageous: in many cases, the very combinations that are missing—such as the application of aggressive treatments to elderly patients—are precisely those of greatest scientific or clinical interest. However, collecting data on such combinations may be impractical, unethical, or infeasible.\\
 \begin{center}
 \begin{tabularx}{\textwidth}{*{2}{>{\centering\arraybackslash}X}} 
\begin{tabular}[b]{cc*{2}{S[table-format=2]}} 
\toprule 
& & \multicolumn{2}{c}{Obesity} \\ 
\cmidrule{3-4} 
\multirow{-2.5}{*}{\makecell{Age}} & \multirow{-2.5}{*}{\makecell{Gender}} & {0} & {1}\\ 
\midrule 
A1 & 0 & 1 & 2 \\
& 1 & 1 & 0 \\\addlinespace
A2 & 0 & 2 & 3 \\
& 1 & 1 & 0 \\\addlinespace
\bottomrule 
\end{tabular}
\captionof{table}{Contingency table obtained from dataset in Table \ref{t0} after collapsing the contingency table by grouping the Age values in ${\rm A1}=\{14\}, {\rm A2}=\{15,16\}$. Here explicitly $Z_{0,1,2}=0$}
\label{t3}
\end{tabularx}
\end{center}
Another possible way to handle generalized random zeros is to ``hide'' the missing entries by collapsing the contingency table. For instance, one could group the Age values as $[{\rm A1},{\rm A2}]=[\{14\} , \{15,16\}]$ so that the contingency table now has no fully-zero rows (or columns) - see table \ref{t3}.  \\
It is worth noting that a general contingency table can, in principle, possess a multitude of generalized random zeros. For instance, we could add to our table \ref{t2} a series of generalized random zeros, simply by considering values of Age outside the variable support defined by the observed data, e.g. ${\rm Age}< 14$ or ${\rm Age} >16$. Those zeroes will be referred to as \emph{spurious generalized random zeros}, because such extra additions are unwarranted and illogical, as they would forcefully add unobserved zeros to the contingency table. This is the reason why we restrict generalized random zeroes to zero entries in the ICT \emph{within the support of the categorical variable under consideration}.\\ 
Now we are ready to formally define generalized random zeros in contingency tables.
\subsection*{Definition of Generalized Incomplete Contingency Table }
Consider a conditional probability $P(Y| X_1,\dots,X_n)$ to be computed on a complete (i.e. with no missing values) dataset $\mathcal{D}$ with categorical variables $X_1,\dots,X_n$ and $y$ with cardinality $I_1,\dots,I_n$ and $J$, respectively. The corresponding $I_1\times\cdots\times I_n \times J$ contingency table $\mathcal{T}$ contains elements  $t_{i_1,\dots,i_n,j}$ with $i_i \in I_i$ and $j \in J$. The set of condition variables ${X_i}$'s are located in the rows of the contingency table, and the outcome $Y$ is the only column.\\
We call $\mathcal{T}$ a generalized incomplete contingency table (GICT) if there exists a combination  $(\bar{i}_1,\dots,\bar{i}_n)$ of values of the variables $X_1,\dots,X_n$ within their support in the dataset $\{i_1\}\times \dots \times \{i_n\}$, for which all cells $t_{\bar{i}_1,\dots,\bar{i}_n,j}=0\;\; \forall j \in J$.\\
\\
It is crucial that the above random zeros appear within the support of all variables, i.e. they are not spurious. In fact, suppose for a moment that we add another value to the variable $X_1$, $i_x \notin I_1$ (see the example above, for ${\rm Age}<14$ or ${\rm Age}>16$): we would have $I_2\times \dots \times I_n$ rows completely filled with random zeros. As we explained earlier, these could simply be deleted from the contingency tables, since they all originate from adding an extra value $i_x$ to $I_1$, which is not present in $\mathcal{D}$ to begin with. Conversely, one can identify the presence of rows of random zeros to be deleted as follows: if $\exists k \in \{1,\dots,N\}$ and a combination of values $(\bar{i}_1,\dots,\bar{i}_{k-1},\bar{i}_{k+1},\dots\bar{i}_n)$  such that $t_{\bar{i}_1,\dots,\bar{i}_{k-1},i_{k},\bar{i}_{k+1},\dots\bar{i}_n,j}=0 \;\;\forall\; i_{k},j\in I_k,J$, then the $I_k$ rows corresponding to the values  $(\bar{i}_1,\dots,\bar{i}_{k-1},\bar{i}_{k+1},\dots,\bar{i}_n)$ in the contingency table can be deleted. Again, it would imply that at least one value among $(\bar{i}_1,\dots,\bar{i}_{k-1},\bar{i}_{k+1},\dots\bar{i}_n)$ sits outside its support $S_{l} \;, l \in N \setminus k$. The procedure can be repeated recursively until all rows of random zeros corresponding to values outside their support have been eliminated. The resulting contingency table may or may not contain generalized random zeros.
Equipped with this definition, we arrive at the formal problem statement: 
\subsection*{Formal Problem Statement}
Let $\mathcal{D}$ be a complete dataset that gives rise to a (set of) generalized incomplete contingency table(s) $\mathcal{T}$ ($\{\mathcal{T}_1, \mathcal{T}_2, \dots \}$) with values of each variable within its support in $\mathcal{D}$ (no spurious generalized zeros). Let $Q$ denote a functional of the joint distributions expressible via $\mathcal{T}$ (e.g., an interventional query or ATE). Define $\mathcal{P}$ as the space of all joint distributions consistent with the observed entries in $\mathcal{T}$. Then, the sharp bounds for $Q$ are:
\[
\min_{P \in \mathcal{P}} Q(P) \le Q^* \le \max_{P \in \mathcal{P}} Q(P)
\]
where $Q^*$ is the true value of the query. Our goal is to characterize $\mathcal{P}$ and compute these extrema numerically. The extremization of queries over unknown cell entries can be interpreted as assuming a non-informative or vacuous prior on the missing counts, akin to adopting a worst/best-case analysis. This is in contrast to techniques that rely on imputations or parametric assumptions over unobserved data.\\

The methodology proposed in this work provides an alternative approach to generalized random zeros which, as far as the authors are aware, has not been considered in the literature, at least not in the fashion proposed here: instead of imputing values, which always requires restrictive and unverifiable assumptions on the statistical models and data generation processes, or deleting the generalized random zeros from the contingency table, we treat them as unknowns and compute, under the reasonable approximation of small frequencies for the random zero cells, the best and worst case impacts they could have on the query. This leads to sharp bounds on the value of interventional queries or Average Treatment Effects (ATEs) under standard approximations. These bounds quantify the uncertainty introduced by generalized random zeros and ensure that, regardless of the true values of the missing cells in the row(s) of a contingency table, the answer to the query for a complete contingency table lies within them.\\
Below, we present an explicit, non-trivial example to describe in detail our approach. 

\section{A simple procedure to evaluate queries in presence of generalized missing data}
Let us consider a slightly different example, where the variable Gender is substituted by the binary variable ``Eating Habits" (H) ( $0=$ Good Habits, $1= $ Bad Habits). The variable Age (A) is considered binary, ``$<40$ years'' ($0$) or ``$>40$ years'' ($1$), as is Obesity (O) ($0=$ No, $1=$ Yes).  The causal graph is shown in Figure \ref{g1} together with a new contingency table, which this time explicitly shows random zeros as 0.
\begin{center}
\begin{tabularx}{\textwidth}{*{2}{>{\centering\arraybackslash}X}}
   \centering
\begin{tikzpicture}
\node[circle] (z) at (2,2) {A};
\node[circle] (x) at (0,0) {H};
\node[circle] (y) at (4,0) {O};
\draw [->,line width=1.5pt] (z) edge (x);
\draw [->,line width=1.5pt] (z) edge (y);
\draw [->,line width=1.5pt] (x) edge (y);
\end{tikzpicture}
\captionof{figure}{Causal graph relating Age (A), Eating Habits (H), Obesity (O) }
\label{g1}
& 
\begin{tabular}[b]{@{}lccc@{}} 
\toprule 
& & \multicolumn{2}{c}{Obesity} \\ 
\cmidrule{3-4} 
\multirow{-2.5}{*}{\makecell{Age}} & \multirow{-2.5}{*}{\makecell{Habits}} & {0} & {1}\\ 
\midrule 
0 & 0 & $0$ & $0$ \\
& 1 & $n_{1}$ & $n_{2}$ \\\addlinespace
1 & 0 & $n_{3}$ & $n_{4}$ \\
& 1 & $0$ & $0$ \\\addlinespace
\bottomrule 
\end{tabular} 
\captionof{table}{Generalized Incomplete contingency table from a dataset generated through the mechanism of causality described by Figure \ref{g1}}
\label{t4}
\end{tabularx}
 \end{center}
 In table \ref{t4}, $n_1,\dots , n_4$ are the numerical entries of the contingency table, with $N=\sum_{i=1}^4 n_i$.\\
 We are interested in answering the simple interventional query $P({\rm O}| \widehat{\rm H})$, with $\widehat{}$  representing an intervention. The result is the well-known ``adjustment formula" \cite{Pearl:2000}:
 \begin{equation}
 \label{eq:adj_formula}
  P({\rm O}| \widehat{\rm H}) = \sum_{{\rm A}=(0,1)} P({\rm O}| {\rm H },{\rm A}) P({\rm A}) \;.  
 \end{equation}
Now, the contingency table \ref{t4} associated with the first factor in eq. \eqref{eq:adj_formula} contains generalized random zeroes, while the table associated with the second factor does not (since it is a marginal probability). So we only need to focus on the analysis of the former factor $P({\rm O}|{\rm H},{\rm A})$ and the corresponding contingency table \ref{t4}.\\
Our proposal to deal with generalized random zeros is quite ingenuous:
\begin{itemize}
    \item Assign $j$ artificial values $x_k^0,\dots x_k^{j-1}$ to each cell in the contingency table  $\mathcal{T}$ that contains a random zero, with $j$ being the multiplicity of the outcome variable in the conditional probability and $k$ indexing the fully-zero rows in $\mathcal{T}$ 
    \item compute the explicit expression for the query, which now depends on the $x_0^i,\dots,x_{k-1}^i$.
    \item find the absolute minima and maxima of the expression, which define the bounds for the value of the query as if the data were not missing but present.
\end{itemize}
Let us work out each step for the example in table \ref{t4}. First, we add unknown counts in the table:
\begin{center}
\begin{tabular}[b]{cc*{2}{S[table-format=2]}} 
\toprule 
& & \multicolumn{2}{c}{Obesity} \\ 
\cmidrule{3-4} 
\multirow{-2.5}{*}{\makecell{Age}} & \multirow{-2.5}{*}{\makecell{Habits}} & {0} & {1}\\ 
\midrule 
0 & 0 & $x_{0}^{0}$ &  $x_{0}^{1}$\\
& 1 & $n_{1}$ & $n_{2}$ \\
\addlinespace
1 & 0 & $n_{3}$ & $n_{4}$ \\
& 1 &  $x_{1}^{0}$ &  $x_{1}^{1}$\\
\addlinespace
\bottomrule 
\end{tabular}
\end{center}
Next we explicitly compute the query \eqref{eq:adj_formula} for the all the values of Obesity and Habit:
\begin{align}
P({\rm O}=0| \widehat{\rm H}=0)&=\frac{x_0^0}{x_0^0+x_0^1} \cdot\frac{n_1+n_2+x_0^0+x_0^1}{N+\sum_{i=0,1} (x_0^i+x_1^i)}+\frac{n_3}{n_3+n_4}\cdot\frac{n_3+n_4+x_1^0+x_1^1}{N+\sum_{i=0,1} (x_0^i+x_1^i)}
\nonumber\\
P({\rm O}=1| \widehat{\rm H}=0)&=\frac{x_0^1}{x_0^0+x_0^1} \cdot\frac{n_1+n_2+x_0^0+x_0^1}{N+\sum_{i=0,1} (x_0^i+x_1^i)}+\frac{n_4}{n_3+n_4}\cdot\frac{n_3+n_4+x_1^0+x_1^1}{N+\sum_{i=0,1} (x_0^i+x_1^i)}
\nonumber\\
P({\rm O}=0| \widehat{\rm H}=1)&=\frac{n_1}{n_1+n_2} \cdot\frac{n_1+n_2+x_0^0+x_0^1}{N+\sum_{i=0,1} (x_0^i+x_1^i)}+\frac{x_1^0}{x_1^0+x_1^1}\cdot\frac{n_3+n_4+x_1^0+x_1^1}{N+\sum_{i=0,1} (x_0^i+x_1^i)}
\nonumber\\
P({\rm O}=1| \widehat{\rm H}=1)&=\frac{n_2}{n_1+n_2} \cdot\frac{n_1+n_2+x_0^0+x_0^1}{N+\sum_{i=0,1} (x_0^i+x_1^i)}+\frac{x_1^1}{x_1^0+x_1^1}\cdot\frac{n_3+n_4+x_1^0+x_1^1}{N+\sum_{i=0,1} (x_0^i+x_1^i)}
\end{align}
Note that the sum over outcomes for each of the two values of the intervention is 1, as expected. \\
Finally, to find the bound, we need to determine the values of $x_0^i,x_1^i$ for which one of the equations is extremized: indeed, in this particular case, the maxima of the first (third) equation correspond to the minima of the second (fourth) equation, and viceversa. \\
In the case where the outcome variable Obesity is not binary, but has $l>2$ values, a linear relationship between the maxima and minima still exists but one needs to extremize first $l-1$ of the $l$ equations for a fixed value of the intervention variable $H$. \\
In this case, the maximum for the first equation is found for $x_0^1=x_1^0=x_1^1=0$ and large values of $x_0^0$ ; the minimum is found for $x_0^0,x_1^0,x_1^1=0$ and large values of $x_0^1$. For these values, the maximum value of the query is 1, and the minimum value is 0. Since these are the natural bounds for an interventional query, in this case, we are not able to bound the result further.

 \subsection{A powerful approximation}

Although the general bounding procedure treats the unknown counts ${x^i_k}$ as free non-negative variables that are potentially unbounded in magnitude, it is often unrealistic to assume that the missing combinations could account for an arbitrarily large proportion of the data. This supports the definition of random zeros as arising from small samples. 
Motivated by this, we consider a small perturbation approximation, where the unknown counts ${x^j_k}$ are assumed to be much smaller than any of the numerical counts $n_i$ in the contingency table. This assumption allows us to expand or simplify the expressions for the queries in the regime where the impact of the missing data is minor but still non-zero. Remarkably, under this approximation, we often obtain significantly tighter bounds, and the resulting expressions become analytically tractable.

In the context of our earlier example, we assume $x^i_0, x^i_1 \ll n_1, n_2, n_3, n_4$, so that the contributions of the missing entries act as a small perturbation to the observed distributions. The four resulting expressions simplify as follows:

\begin{align}
P({\rm O}=0| \widehat{\rm H}=0)&= \pi_0\cdot\frac{n_1+n_2}{N}+\frac{n_3}{n_3+n_4}\cdot\frac{n_3+n_4}{N}
\nonumber\\
P({\rm O}=1| \widehat{\rm H}=0)&=(1-\pi_0) \cdot\frac{n_1+n_2}{N}+\frac{n_4}{n_3+n_4}\cdot\frac{n_3+n_4}{N}
\nonumber\\
P({\rm O}=0| \widehat{\rm H}=1)&=\frac{n_1}{n_1+n_2} \cdot\frac{n_1+n_2}{N}+\pi_1\cdot\frac{n_3+n_4}{N}
\nonumber\\
P({\rm O}=1| \widehat{\rm H}=1)&=\frac{n_2}{n_1+n_2} \cdot\frac{n_1+n_2}{N}+(1-\pi_1)\cdot\frac{n_3+n_4}{N}
k\end{align}
where $\pi_0 = \frac{x^0_0}{x^0_0+x_0^1}$ and $\pi_1=\frac{x_1^0}{x^0_1+x^1_1}$. It is easy to see that the maximum (minimum) of the first equation corresponds to the value $\pi_0=1$ ($\pi_0=0$), and viceversa for the second equation. Similarly, one can maximize (minimize) the last two queries and obtain similar results. Again, the maxima and minima for the query are obtained for extreme values of the probabilities, $\pi_0, \pi_1$ ( $0$ or $1$), even though this time they are given by:
\begin{align}
    P({\rm O}=0| \widehat{\rm H}=0)_{\rm Max}&= \frac{n_1+n_2+n_3}{N} <1 \\
    \nonumber\\
    P({\rm O}=0| \widehat{\rm H}=0)_{\rm Min}&= \frac{n_3}{N} >0
\end{align} 
If the query contains multiple conditional probabilities, the expression will always be linear in each $\pi$, but there may be products of $\pi$'s which will also change their values at maximum/minimum.\\

As we can see, the expressions for queries are typically linear or ratio-of-linear functions of the unknowns $x_k^i$. In such cases, the minimization and maximization of the query can be cast as a convex program over a bounded domain (e.g., $x^i_k \ge 0$, $\sum_{i,k} x^i_k \le \epsilon$ for small $\epsilon$). This enables efficient computation of bounds via convex optimization techniques. We outline the high-level steps in Algorithm ~\ref{alg:bounding} below.
\begin{algorithm}[h!]
\caption{Bounding Interventional Queries with Generalized Missingness}
\label{alg:bounding}
\begin{flushleft}
\textbf{Input:} target query $Q$; dataset $\mathcal{D}$ with missing entries\\
\textbf{Output:} Sharp bounds $[\min Q, \max Q]$
\begin{enumerate}
    \item Eliminate missing entries in $\mathcal{D}$ by imputation or deletion of corresponding row(s)
    \item for each conditional probability $P_i(Y|X_1,\dots,X_n)$ inside $Q$, create an associated contingency table $\mathcal{T}_i$ from the dataset $\mathcal{D}$
    \begin{enumerate}
    \item run Algorithm \ref{alg:clean} on $\mathcal{T}_i$ to eliminate spurious generalized zeros
    \item Identify the $k$ rows of generalized random zeros 
    \item For each generalized zero corresponding to one of the values of the outcome variable $Y$ of cardinality $J$, introduce one variable $\pi_k^l=\frac{x_k^l}{\sum_j x_k^j}$ with $l \in \{0,\dots,J-1\}$. This definition satisfies the basic probability constraint $\sum_l \pi_k^l =1 \;\; \forall\; k $; hence for each fixed $k$, only $j-1$ of these probabilities $\pi^l_k$ are independent
    \end{enumerate}
    \item Express the query $Q$ symbolically in terms of both observed counts $n_l$ and the variables $\pi^j_k$'s, under the simplifying assumption that $x^j_k \ll n_l $
    \item Extremize $Q$ over all $\pi_k$'s subject to non-negativity and normalization constraints (e.g., rows of probabilities sum to one).
\end{enumerate}
\end{flushleft}
\end{algorithm}
\begin{algorithm}[h!]
\caption{Delete spurious generalized random zeros in ICT}
\label{alg:clean}
\begin{flushleft}
\textbf{Input:} A $I_1\times\dots\times I_n \times J$ contingency table $\mathcal{T}$ with (spurious) generalized random zeroes\\
\textbf{Output:} $\mathcal{T}$ with values outside their support in $\mathcal{D}$ deleted
\begin{enumerate}
    \item Check if there exists a $r \in \{1,\dots n\}$ and a combination of values $\bar{i}_1,\dots,\bar{i}_{r-1},\bar{i}_{r+1},\dots \bar{i}_N$ such that $t_{\bar{i}_1,\dots,\bar{i}_{r-1},i_r,\bar{i}_{r+1},\dots \bar{i}_n,j}=0$ $\forall\; (i_r,j) \in (I_r,J)$
    \item Deleted all the rows indexed by $\bar{i}_1,\dots,\bar{i}_{r-1},\bar{i}_{r+1},\dots \bar{i}_n$, $\forall\; (i_r,j) \in (I_r,J)$
    \item Call back Algorithm \ref{alg:clean} until all spurious zeros have been deleted
\end{enumerate}
\end{flushleft}
\end{algorithm}
\newpage
\section{Conclusion}

We have introduced a first principles approach for handling incomplete contingency tables, where certain combinations of variables are completely unobserved. In such cases, traditional methods like maximum likelihood estimation or multiple imputation are either undefined or require sometimes-unverifiable assumptions about the missing data mechanism. 

Instead of imputing or discarding missing entries, we propose a nonparametric bounding method by treating the unknown counts as free parameters and deriving symbolic expressions for interventional queries in terms of these unknowns. By extremizing the resulting expressions subject to the natural constraints of probability theory, we obtain bounds within which the true query value $ Q^* $ must lie, assuming only the observed data and logical consistency.

This approach can be interpreted as a robust inference method under weak constraints on the low-frequency of random zeros entries in the contingency table. It avoids unwarranted assumptions and guarantees that the true value of the causal query, had the full table been available, lies within the derived bounds. While these bounds can be wide in some cases, for example, when large regions of the table are unobserved, they still quantify the uncertainty introduced by generalized zeros in a transparent way.

Moreover, we showed in an illustrative example how simply restricting the missing entries in the contingency to be small relative to the observed counts can lead to significantly narrower and more informative bounds. This offers a practical compromise between conservatism and informativeness, suitable for real-world datasets such as medical data, where generalized random zeros  are very common due to ethical, demographic and design constraints as well as small sample sizes.


\begin{thebibliography}{99}
\bibitem{Kang:13}
H. Kang, ``The prevention and handling of the missing data", Korean J Anesthesiol. 64(5):402-6, (2013) doi: 10.4097/kjae.2013.64.5.402. Epub 2013 May 24. PMID: 23741561; PMCID: PMC3668100.
\bibitem{review2}
K. Sathishkumar, R. Jaisankar and T. Ramamoorthy, ``A review of statistical methods for addressing missing data", Advanced Applications in Statistics, {\bf 92}, (8), (2025), https://pphmjopenaccess.com/aas/article/view/2950
\bibitem{Pearl:2020}
K. Mohan and J. Pearl, ``Graphical Models for Processing Missing Data",Journal of the American Statistical Association, {\bf 116}, 534 (2021), https://doi.org/10.1080/01621459.2021.1874961
\bibitem{Baker:85}
R. J. Baker, M. R. B. Clarke and  P. W. Lane, ``Zero entries in contingency tables", Comput. Stat. Data Anal., {\bf 3} (33-45), (1985), https://doi.org/10.1016/0167-9473(85)90056-8 
\bibitem{smith:90}
W. B. Smith and A. R. Parsa, `` Contingency tables with structural zeros",
Comm. in Stat. - Theory and Methods, {\bf 19} (12) (1990), https://doi.org/10.1080/03610929008830453
\bibitem{Dempster:18}
A. P. Dempster, N. M. Laird and D. B. Rubin, ``Maximum Likelihood from Incomplete Data Via the EM Algorithm", Journal of the Royal Statistical Society: Series B (Methodological), {\bf 39} (1), (2018), https://doi.org/10.1111/j.2517-6161.1977.tb01600.x
\bibitem{brown:83}
M. B. Brown and C. Fuchs, ``On maximum likelihood estimation in sparse contingency tables", Comput. Stat. Data Anal., {\bf 1} (3-15), (1983), https://doi.org/10.1016/0167-9473(83)90059-2
\bibitem{Brzezińska:15}
J. Brzezińska, ``The Problem of Zero Cells in the Analysis of Contingency Tables", Zeszyty Naukowe Uniwersytetu Ekonomicznego w Krakowie, 49-61, (2015),  10.15678/ZNUEK.2015.0941.0504
\bibitem{fienberg:18}
S. E. Fienberg, ``The Analysis of Cross-Classified Categorical Data", Royal Statistical Society. Journal. Series A: General, {\bf 141}, (4), (2018), https://doi.org/10.2307/2344495 
\bibitem{Consonni:07}
G. Consonni and G. Pistone, ``Algebraic Bayesian analysis of contingency tables with possibly zero-probability cells", Statistica Sinica {\bf 17} (4), (2007),  https://doi.org/10.48550/arXiv.math/0703123
\bibitem{Pearl:2000}
J. Pearl , ``Causality: Models, Reasoning, and Inference'', Cambridge University Press, 2000, ISBN: 0-521-77362-8 
\end{thebibliography}
\end{document}